%% file: straub.tex
\documentclass[12pt]{article}
\usepackage{graphicx}
\usepackage{amsmath}
\newcommand\pubnumber{}
\newcommand\pubdate{\today}

\def\Title#1{\begin{center} {\Large #1 } \end{center}}
\def\Author#1{\begin{center}{ \sc #1} \end{center}}
\def\Address#1{\begin{center}{ \it #1} \end{center}}

\newcommand\pubblock{\rightline{\begin{tabular}{l} \pubnumber\\
         \pubdate  \end{tabular}}}
\newenvironment{Abstract}{\begin{quotation}}{\end{quotation}}
\newenvironment{Presented}{\begin{quotation} \begin{center} 
             PRESENTED AT\end{center}\bigskip 
      \begin{center}\begin{large}}{\end{large}\end{center} \end{quotation}}
\def\Acknowledgements{\bigskip  \bigskip \begin{center} \begin{large}
             \bf ACKNOWLEDGEMENTS \end{large}\end{center}}

\input econfmacros.tex

\begin{document}
\begin{titlepage}
\pubblock

\vfill
\Title{New physics correlations in rare decays}
\vfill
\Author{David M. Straub}
\Address{%
Physik-Department, Technische Universit\"at M\"unchen, 85748 Garching, Germany\\
and\\
Scuola Normale Superiore and INFN, Piazza dei Cavalieri 7, 56126 Pisa, Italy}
\vfill
\begin{Abstract}
This talk gives an overview of correlations between new physics effects in rare $B$ and $K$ decays arising in concrete models or on a model-independent basis and discusses how these correlations can be utilized to distinguish between different classes of new physics scenarios.
\end{Abstract}
\vfill
\begin{Presented}
6th International Workshop on the CKM Unitarity Triangle \\
Warwick, UK, September 6--10, 2010
\end{Presented}
\vfill
\end{titlepage}
\def\thefootnote{\fnsymbol{footnote}}
\setcounter{footnote}{0}

\section{Introduction: Rare decays in the LHC era}

The search for rare decays is as promising as it is challenging. The reason for the rareness of flavour-changing neutral current (FCNC) decays in the Standard Model (SM) is that they arise only at the loop level by means of the GIM mechanism and can be further suppressed by the hierarchies in the CKM matrix or helicity suppression. In models of new physics (NP), any of these suppression mechanisms can be lifted, leading to potentially large enhancement of the SM rates. This sensitivity to short-distance physics is what makes rare FCNC decays such valuable probes of physics beyond the SM.

To date, no significant deviation of rare decay observables from their SM expectations has been found. Still, in many decays there is plenty of room for NP contributions, and a number of recently started or planned experiments will put this under scrutiny. These experiments include LHCb, ATLAS and CMS at LHC, Belle-II at the upgraded KEKB, the SuperB project, NA62 at CERN and KOTO at JPARC.

Now that the high-energy frontier is being probed by LHC, the role of rare decays is shifted: rather than establishing a signal of NP, rare decays are more likely to play an important role in distinguishing different models of NP with similar LHC signatures, or to probe the flavour structure of the NP theory. Indeed, solving the ``flavour puzzle'' is unlikely to be achieved on the basis of collider data only. However, to fully harness the power of rare decays as a probe of NP, it is crucial to exploit correlations between observables and the implications for NP scenarios.

In this talk, I will focus on rare radiative and semileptonic $B$ and $K$ decays.
An (incomplete) list of interesting rare decays with a $b\to s$, $b\to d$ or $s\to d$ transition and a charged lepton pair, neutrino pair or photon in the final state is given in table~\ref{tab:obs}.

\begin{table}[tbp]
 \centering
\renewcommand{\tabcolsep}{0.5cm}
\renewcommand{\arraystretch}{1.2}
\begin{tabular}{|c|c|c|c|}
\hline
 & \boldmath $\nu\bar\nu$ & \boldmath $\ell^+\ell^-$ & \boldmath $\gamma$\\
\hline
\boldmath $b\to s$ & $B\to X_s\nu\bar\nu$ & $B\to X_s\ell^+\ell^-$ & $B\to X_s\gamma$ \\
$(\sim\lambda^2)$      & $B\to K^*\nu\bar\nu$ & $B\to K^*\ell^+\ell^-$ & $B\to K^*\gamma$ \\
         & $B\to K\nu\bar\nu$ & $B\to K\ell^+\ell^-$ & $B\to K\gamma$ \\
         & & $B_s\to\mu^+\mu^-$ & \\
\hline
\boldmath $b\to d$ & $B\to X_d\nu\bar\nu$ & $B\to X_d\ell^+\ell^-$ & $B\to X_d\gamma$ \\
$(\sim\lambda^3)$   & & $B_d\to\mu^+\mu^-$ &  \\
\hline
\boldmath $s\to d$ & $K_L\to\pi^0\nu\bar\nu$ & $K_L\to\pi^0\ell^+\ell^-$ & \\
$(\sim\lambda^5)$     & $K^+\to\pi^+\nu\bar\nu$ & $K_L\to\ell^+\ell^-$ & \\
\hline
\end{tabular}
\renewcommand{\arraystretch}{1.0}
 \caption{(Incomplete) list of theoretically interesting and experimentally promising rare semileptonic and radiative $B$ and $K$ decays. The approximate CKM suppression within the SM is indicated in the leftmost column in terms of the Wolfenstein parameter $\lambda$.}
 \label{tab:obs}
\end{table}

\section{Correlations between rare decays}
\label{sec:corr}

In all models considered in this talk, NP enters the observables via the Wilson coefficients of the operators in the effective Hamiltonian
\begin{equation}
\mathcal H_\text{eff}^{d_i\to d_j} = -\frac{4G_F}{\sqrt{2}}V_{ti}V_{tj}^*\sum_k(C_k^{ij}Q_k^{ij}+C_k^{\prime ij}Q_k^{\prime ij})
\,,
\label{eq:heff}
\end{equation}
\begin{align}
Q_{7}^{(\prime)ij} &= 
m_b
(\bar d_j \sigma_{\mu \nu} P_{R,L} d_i) F^{\mu \nu} ,&
Q_{8}^{(\prime)ij} &= 
m_b
(\bar d_j \sigma_{\mu \nu} T^a P_{R,L} d_i) G^{\mu \nu \, a} ,&
\nonumber\\
Q_{9}^{(\prime)ij} &= 
(\bar d_j \gamma_{\mu} P_{L,R} d_i)(\bar{\ell} \gamma^\mu \ell) ,&
Q_{10}^{(\prime)ij} &= 
(\bar d_j  \gamma_{\mu} P_{L,R} d_i)(  \bar{\ell} \gamma^\mu \gamma_5 \ell) ,&
\nonumber\\
Q_{S}^{(\prime)ij} &= 
m_b (\bar d_j P_{R,L} d_i)(  \bar{\ell} \ell) ,&
Q_{P}^{(\prime)ij} &= 
m_b (\bar d_j P_{R,L} d_i)(  \bar{\ell} \gamma_5 \ell) ,&
\nonumber\\
Q_{L,R}^{ij} &= 
(\bar d_j \gamma_\mu P_{L,R} d_i)(  \bar{\nu} \gamma^\mu P_L \nu) .&
\label{eq:ops}
\end{align}
In its full generality, this effective theory does not allow to draw any conclusions on correlations between rare decays, with the exception of consistency conditions like e.g. the Grossman-Nir bound, limiting the ratio $\text{BR}(K_L\to\pi^0\nu\bar\nu)/\text{BR}(K^+\to\pi^+\nu\bar\nu)$ \cite{Grossman:1997sk}. To obtain correlations between the predictions for various observables, assumptions have to be made on the NP theory.

There are basically three strategies to study such correlations.

\begin{enumerate}
\item {\bf Dominance of certain operators or topologies}
\\
Assuming only a limited subset of the operators in (\ref{eq:ops}) to be modified with respect to the SM, correlations can arise between observables sensitive to the respective operators\footnote{%
Of course, restricting to the operators in (\ref{eq:ops}) is already an assumption of this type.
}. An experimentally observed deviation from these correlations would then unambiguously imply the presence of additional operators.
Likewise, assuming NP to enter dominantly through certain topologies, e.g. only through $Z$ penguins, correlations between different operators can arise.
In both cases, correlations can arise between
decays with charged leptons, neutrinos or a photon in the final state, but only with the same (quark) flavour content.
Thus, this strategy leads to {\em horizontal correlations} between the observables in table~\ref{tab:obs}.
\item {\bf Flavour symmetries}
\\
Flavour symmetries allow to predict correlations among observables differing {\em only} by their flavour content, i.e. correlations between $b\to s$, $b\to d$ and $s\to d$ transitions. The most powerful symmetry principle is the assumption of Minimal Flavour Violation (MFV) \cite{D'Ambrosio:2002ex}. In the MFV case, flavour violation is governed solely by the CKM matrix, such that all the Wilson coefficients in (\ref{eq:heff}) are flavour independent:
\begin{equation}
C_i^{(\prime)bs} = C_i^{(\prime)bd} = C_i^{(\prime)sd} \,.
\label{eq:flavind}
\end{equation}
This strategy then leads to {\em vertical correlations} between the observables in table~\ref{tab:obs}.
For a review of correlations in MFV, see \cite{Hurth:2008jc}.
\item {\bf Concrete NP models}
\\
In concrete NP models, correlations between observables can arise if the NP contributions to the Wilson coefficients depend on a small number of parameters or if the parameters they depend on are constrained by other observables or theoretical considerations. These correlations can be vastly different from model to model and so allow in many cases to discriminate between different NP models solely on the basis of rare decay correlations. For a recent review of this strategy, see \cite{Buras:2010wr}.
\end{enumerate}

Naturally, one can also combine these strategies. The class of models fulfilling constrained MFV (CMFV, \cite{Buras:2000dm}) correspond to strategy 2.\ with the additional assumption that only the operators present in the SM are relevant, corresponding to strategy 1. Concrete NP models also frequently fall into category 2.\ (e.g. by exhibiting MFV) or into category 1.\ (e.g. predicting the absence of right-handed currents).

The rest of the talk will focus on the confrontation of strategies 2.\ and 3.\ in the case of $s\to d\nu\bar\nu$ and $b\to (s,d)\ell^+\ell^-$ decays, and on strategy 1.\ in the case of $b\to s\ell^+\ell^-$ and $b\to s\nu\bar\nu$ decays.

\section{Correlations in MFV and concrete NP models}

\subsection{\boldmath $K_L\to\pi^0\nu\bar\nu$ vs. $K^+\to\pi^+\nu\bar\nu$}

\begin{figure}[tbp]
 \centering
 \includegraphics[width=10.5cm]{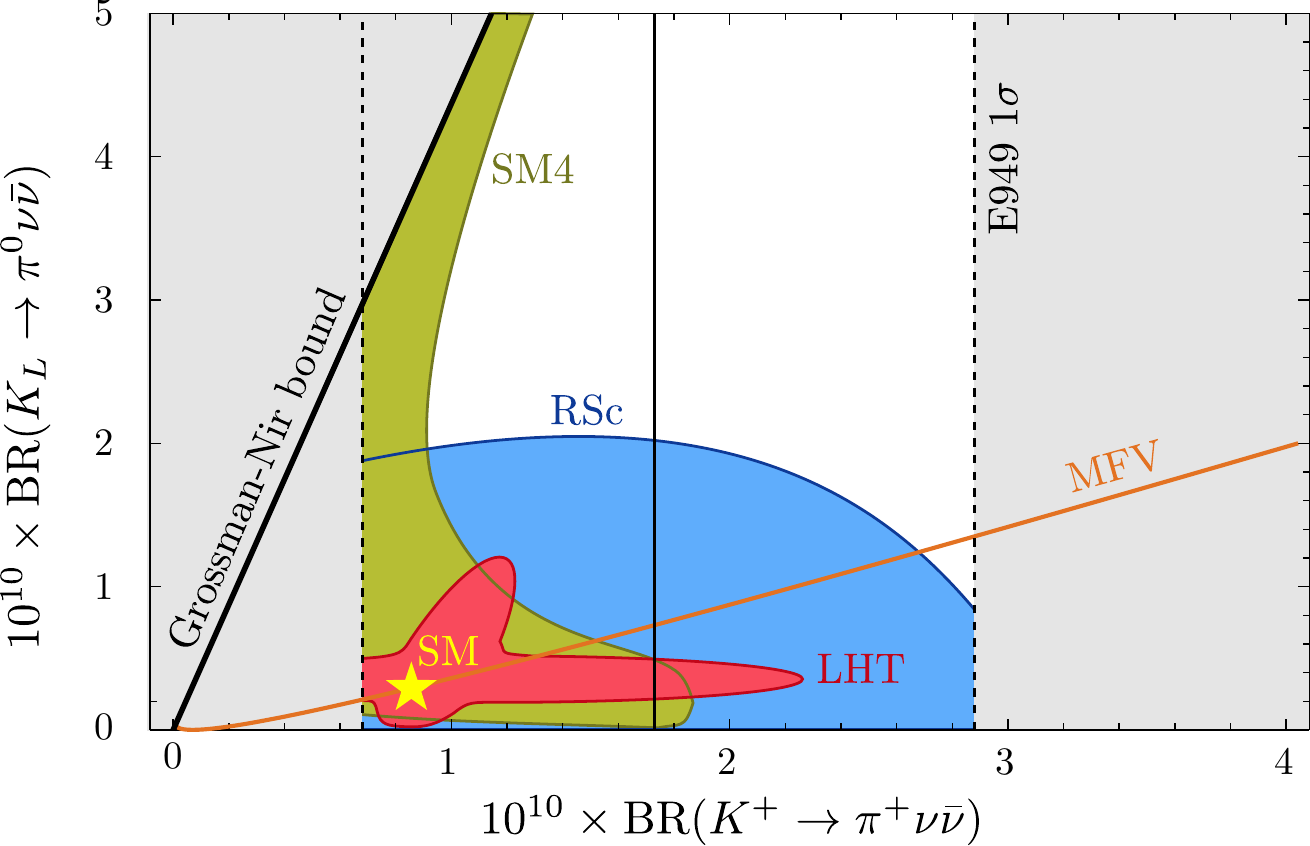}
 \caption{Correlation between the branching ratios of $K_L\to\pi^0\nu\bar\nu$ and $K^+\to\pi^+\nu\bar\nu$ in MFV and three concrete NP models. The gray area is ruled out experimentally or model-independently by the GN bound. The SM point is marked by a star.}
 \label{fig:kpinunu}
\end{figure}

The rare decays $K_L\to\pi^0\nu\bar\nu$ and $K^+\to\pi^+\nu\bar\nu$ are the golden modes of $K$ physics; they are extremely rare in the SM due to a power-like GIM suppression and theoretically clean since the form factors can be extracted from experimental data on $K_{\ell3}$ decays. Their branching ratios can be written as\footnote{%
Assuming, for simplicity, that the Wilson coefficients do not depend on the neutrino flavour.
}
\begin{align}
\text{BR}(K^+\to\pi^+\nu\bar\nu) &= \kappa_+ \;\left| \xi X - P_{(u,c)} \right|^2
\,,&
\text{BR}(K_L\to\pi^0\nu\bar\nu) &= \kappa_L ~\text{Im}\!\left(\xi X\right)^2
\,,
\end{align}
where $\xi={V_{ts}^*V_{tb}}/{V_{us}^5}$ and $X=-\frac{16\pi^2}{e^2s_w^2}(C_L^{sd}+C_R^{sd})$ in the notation of eq.~(\ref{eq:ops}). In every model beyond the SM, both branching ratios only depend on one complex quantity $X$. The model-independent Grossman-Nir bound shown in figure~\ref{fig:kpinunu} then simply arises from the fact that the imaginary part of a complex number has to be smaller than or equal to its absolute value, corrected by the lifetime differences and isospin breaking effects.

In MFV, assuming that there is no CP violation beyond the CKM phase, the contribution to both branching ratios is simply given by a real contribution to $X$, leading to a clean correlation shown in figure~\ref{fig:kpinunu} as an orange line\footnote{%
It should be stressed that this correlation is simply a consequence of assuming no new sources of CP violation, not of a flavour symmetry in the sense of strategy 2.\ above.
}.

In concrete NP models, spectacular deviations from the MFV prediction are possible.
The coloured areas in figure~\ref{fig:kpinunu} show the correlations arising in three models: the Littlest Higgs model with T-parity (LHT, \cite{Blanke:2009am}), the Randall-Sundrum model with custodial protection (RSc, \cite{Blanke:2008yr}) and the Standard Model with a sequential fourth generation (SM4, \cite{Buras:2010pi})\footnote{%
Since parametrization choices, scan statistics etc. make the comparison between predictions in different models quite delicate, the regions indicated figures~\ref{fig:kpinunu} and \ref{fig:bqmumu} are to be undestood as merely indicative. For the actual results, one should consult the original publications.
}.
In the RSc model, there is no correlation, but the overall effects are limited. In the LHT, mild enhancements of both branching ratios are possible, but not simultaneously. In the SM4, the GN bound can be saturated even for large values of $\text{BR}(K^+\to\pi^+\nu\bar\nu)$. In the MSSM with large flavour violating trilinear terms in the up-squark sector, large (uncorrelated) effects are possible in both decays \cite{Isidori:2006qy}.

\subsection{\boldmath $B_s\to\mu^+\mu^-$ vs. $B_d\to\mu^+\mu^-$}

The correlation between the decays $B_s\to\mu^+\mu^-$ and $B_d\to\mu^+\mu^-$ is an example of a ``vertical'' correlation mentioned in section~\ref{sec:corr}. Beyond the SM, their branching ratios can be written as
\begin{equation}
 \text{BR}(B_q\to\mu^+\mu^-) \propto |S|^2 \left( 1 - 4x_\mu^2 \right) + |P|^2,
\label{eq:BRBqmumu}
\end{equation}
\begin{equation}
S =  C_S^{bq} - C_S^{\prime bq}
\,, \qquad
P = C_P^{bq} - C_P^{\prime bq} + 2 x_\mu (C_{10}^{bq} - C_{10}^{\prime bq})
\,, \qquad
x_\mu=m_\mu/m_{B_s}
\,.
\label{eq:BRBsmumu2}
\end{equation}
Order-of-magnitude enhancements of these branching ratios are only possible in the presence of sizable contributions from scalar or pseudoscalar operators. In two-Higgs-doublet models, the contribution to $C_S^{bq}$ from neutral Higgs exchange scales as $\tan\beta^2$, where $\tan\beta$ is the ratio of the two Higgs VEVs. In the MSSM, the non-holomorphic corrections to the Yukawa couplings even enhance this contribution to $\tan\beta^3$.

\begin{figure}[tbp]
 \centering
 \includegraphics[width=10.5cm]{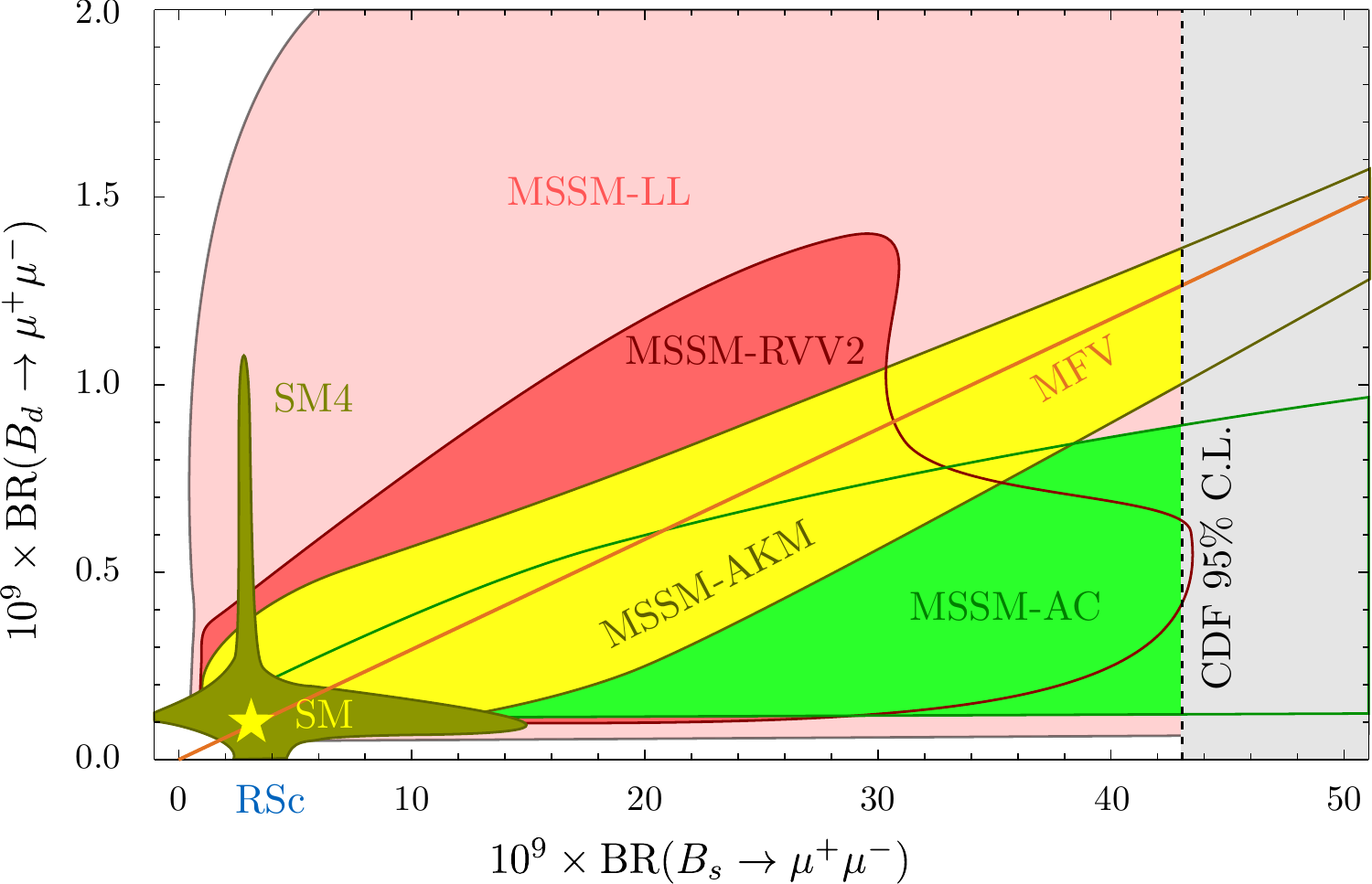}
 \caption{Correlation between the branching ratios of $B_s\to\mu^+\mu^-$ and $B_d\to\mu^+\mu^-$ in MFV, the SM4 and four SUSY flavour models. The gray area is ruled out experimentally. The SM point is marked by a star.}
 \label{fig:bqmumu}
\end{figure}

Figure~\ref{fig:bqmumu} shows the correlation between $\text{BR}(B_s\to\mu^+\mu^-)$ and $\text{BR}(B_d\to\mu^+\mu^-)$ in MFV, the SM4 and four SUSY flavour models\footnote{%
The acronyms stand for the models by Agashe and Carone (AC, \cite{Agashe:2003rj}), Ross, Velasco-Sevilla and Vives (RVV2, \cite{Ross:2004qn}), Antusch, King and Malinsky (AKM, \cite{Antusch:2007re}) and a model with left-handed currents only (LL, \cite{Hall:1995es}).

}
analyzed in detail in \cite{Altmannshofer:2009ne}. The MFV line, shown in orange, is obtained from the flavour independence of the Wilson coefficients, cf. eq. (\ref{eq:flavind}). The largest effects are obtained in the SUSY flavour models due to the above-mentioned Higgs-mediated contributions. While in some models, like the one with left-handed currents only, there is no correlation, others, like the AKM model, stay close to the MFV prediction. The AC model predicts that $\text{BR}(B_s\to\mu^+\mu^-)$ is always larger than the value implied by MFV. In the RSc and LHT models, the effects in both observables are below current and projected experimental sensitivities. In the SM with four generations, on the other hand, both branching ratios can be visibly enhanced inspite of the absence of scalar operators, but there is a strict anticorrelation. An observed simultaneous enhancement would thus immediately rule out the SM4.

\section{Correlations from dominating operators}

\subsection{\boldmath $B\to X_s\gamma$ vs. $B\to K^*\mu^+\mu^-$}

$B\to K^*(\to K\pi)\mu^+\mu^-$ gives access to a large number of observables sensitive to new physics via its angular distribution. While the branching ratio itself has a considerable theoretical uncertainty due to the hadronic form factors, angular observables normalized to the branching ratio are theoretically much cleaner. To study correlations among these observables and between $B\to X_s\gamma$ and $B\to K^*\mu^+\mu^-$, let us follow strategy 1.\ and assume new physics to affect only the magnetic penguin operator $Q_7$ and its chirality-flipped counterpart, $Q_7'$. (This also covers models where NP enters $Q_8^{(\prime)}$, since, to a good approximation, $C_7^{(\prime)}$ and $C_8^{(\prime)}$ enter the observables in a fixed linear combination).
There are several concrete NP models where such a situation is realized. For example, the MSSM with MFV and flavour blind phases (FBMSSM) leads mostly to complex contributions to $C_7$ (and $C_8$) \cite{Altmannshofer:2008hc,Paradisi:2009ey,Altmannshofer:2009ne}. Complex contributions to $C_{7,8}'$ can be generated e.g. in the MSSM with a large complex 32-element in the down-type squark trilinear coupling matrix \cite{Altmannshofer:2008dz}.

Under this assumption, the most important effects in $B\to K^*\mu^+\mu^-$ are generated in the CP-averaged angular coefficients $S_4$, $S_5$ and $S_6^s$ and in the CP asymmetries $A_7$, $A_8$ and $A_9$, in the notation of \cite{Altmannshofer:2008dz}. The most important constraint arises from the $B\to X_s\gamma$ branching ratio. $S_6^s$ is proportional to the well-known forward-backward asymmetry. $A_{7,8,9}$ are the T-odd CP asymmetries, which are not suppressed by small strong phases \cite{Bobeth:2008ij}.

\begin{figure}[tbp]
 \centering
 \includegraphics[width=0.79\textwidth]{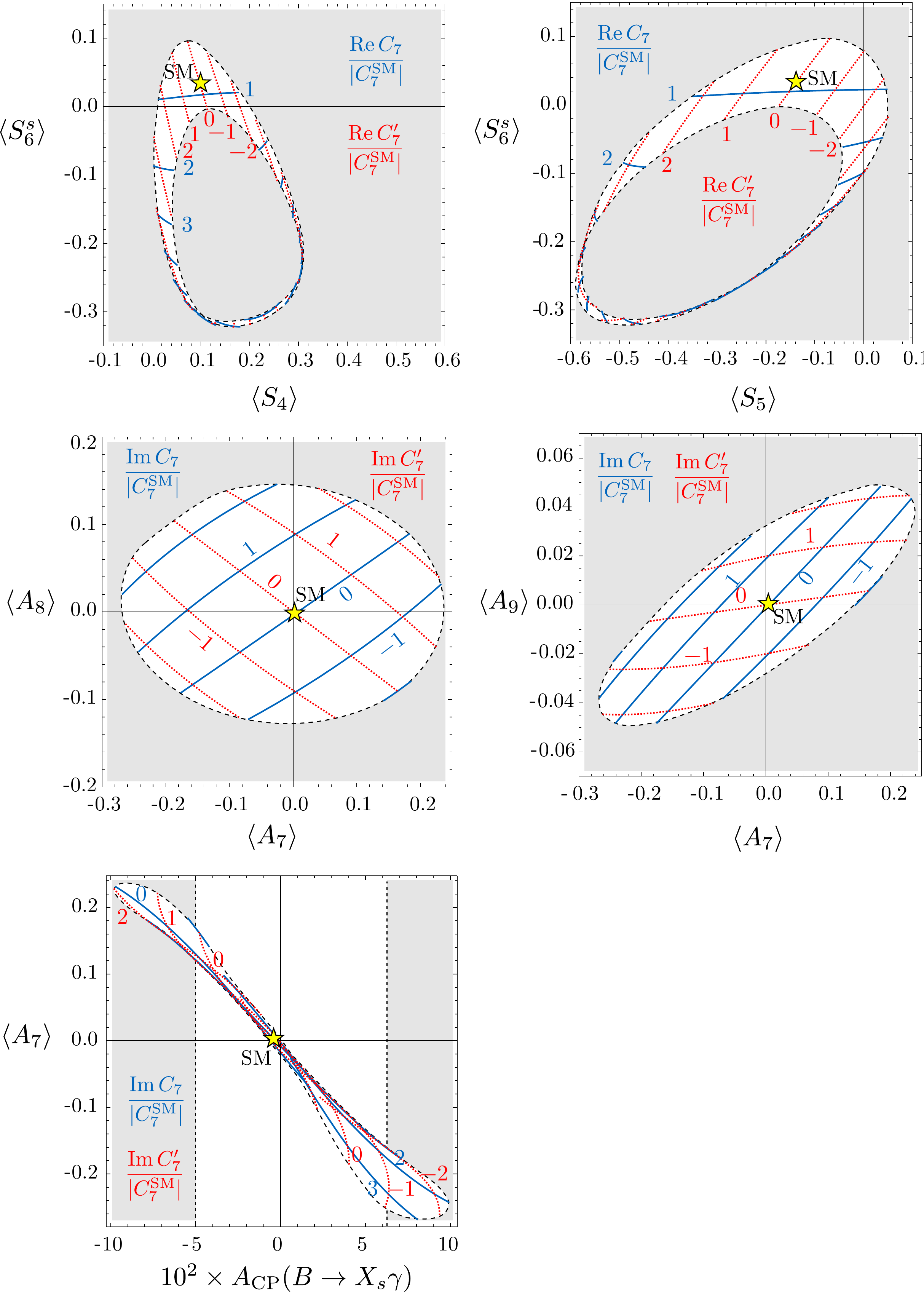}
 \caption{Correlation between the $B\to K^*\mu^+\mu^-$ CP-averaged angular coefficients (first row), $B\to K^*\mu^+\mu^-$ CP asymmetries (second row) and between one of the CP asymmetries in $B\to K^*\mu^+\mu^-$ and the one in $B\to X_s\gamma$. All correlations are based on the assumption that NP enters only the magnetic penguin operators. The contours indicate the real part (first row) or imaginary part (second and third rows) of $C_7$ (blue contours) and $C_7'$ (dotted red contours) at a scale of 160~GeV, normalized to the SM value of $C_7$.
The SM point is marked by a star.}
 \label{fig:BKsmumu}
\end{figure}

The two plots in the first row of figure~\ref{fig:BKsmumu} show the correlation between $S_4$, $S_5$ and $S_6^s$ integrated in the low dilepton invariant mass region, $q^2\in[1,6]\,\text{GeV}^2$.
The gray area is disfavoured at more than $2\sigma$ by the $\text{BR}(B\to X_s\gamma)$ constraint.
The contours show the value of the real parts of $C_7$ (blue contours) and $C_7'$ (red dotted contours) at the scale of 160~GeV, normalized to the SM value of $C_7$ ($C_7'$ is close to zero in the SM). The Wilson coefficients' imaginary parts play next to no role, since these are the CP averaged observables.

The two plots in the second row of figure~\ref{fig:BKsmumu} show the correlation between the $q^2$-integrated $A_7$, $A_8$ and $A_9$ in the considered scenario, after imposing the bound from  $\text{BR}(B\to X_s\gamma)$. In these cases, the contours show the {\em imaginary} parts of $C_7$ and $C_7'$, normalized to the absolute value of $C_7^\text{SM}$.

The remarkable feature of the correlations shown in these four plots is that a combined analysis of these abservables allows, assuming the dominance of magnetic penguins, to infer the magnitude, phase, and chirality structure of the Wilson coefficients. This is in contrast to the $B\to X_s\gamma$ branching ratio, which only gives access to a single quantity. As an example, in models with complex $C_7$ like the FBMSSM mentioned above, positive $\langle A_7 \rangle$ always implies negative $\langle A_8 \rangle$; in models with right-handed currents, it implies positive $\langle A_8 \rangle$.

Finally, the plot in the last row of figure~\ref{fig:BKsmumu} shows the correlation between $\langle A_7 \rangle$ and the direct CP asymmetry in $B\to X_s\gamma$ that arises in this framework. This is an example of a ``horizontal'' correlation arising from strategy 1.

\subsection{\boldmath $B\to K\nu\bar\nu$ vs. $B\to K^*\nu\bar\nu$}

$B^+\to K^+\nu\bar\nu$ is the analogue of $K^+\to \pi^+\nu\bar\nu$ in the $b\to s$ sector and is sensitive to $|C_L^{bs}+C_R^{bs}|^2$. While there exists no analogue of the $K_L$ decay and thus no direct probe of CP violation in $b\to s\nu\bar\nu$ transitions, there is another exclusive decay, $B\to K^*\nu\bar\nu$, which gives access to the branching ratio and the $K^*$ longitudinal polarization fraction $F_L$ (via the angular analysis of $K^*\to K\pi$), as well as the inclusive decay $B\to X_s\nu\bar\nu$. All four observables depend in a different way on the two Wilson coefficients $C_{L,R}^{bs}$, but can be expressed in terms of two real combinations of them,
\begin{equation}  \label{eq:epsetadef}
 \epsilon = \frac{\sqrt{ |C^{bs}_L|^2 + |C^{bs}_R|^2}}{|(C^{bs}_L)^\text{SM}|}~, \qquad
 \eta = \frac{-\text{Re}\left(C^{bs}_L C_R^{bs *}\right)}{|C^{bs}_L|^2 + |C^{bs}_R|^2}~.
\end{equation}
In particular,
\begin{align}
\label{eq:epseta-BKsnn}
 \text{BR}(B \to K^* \nu\bar\nu) & = \text{BR}(B \to K^* \nu\bar\nu)_\text{SM} \times (1 + 1.31 \,\eta)\epsilon^2~, \\
\label{eq:epseta-BKnn}
 \text{BR}(B \to K \nu\bar\nu)   & = \text{BR}(B \to K^* \nu\bar\nu)_\text{SM} \times (1 - 2\,\eta)\epsilon^2~.
\end{align}

\begin{figure}[tbp]
 \centering
 \includegraphics[width=10.5cm]{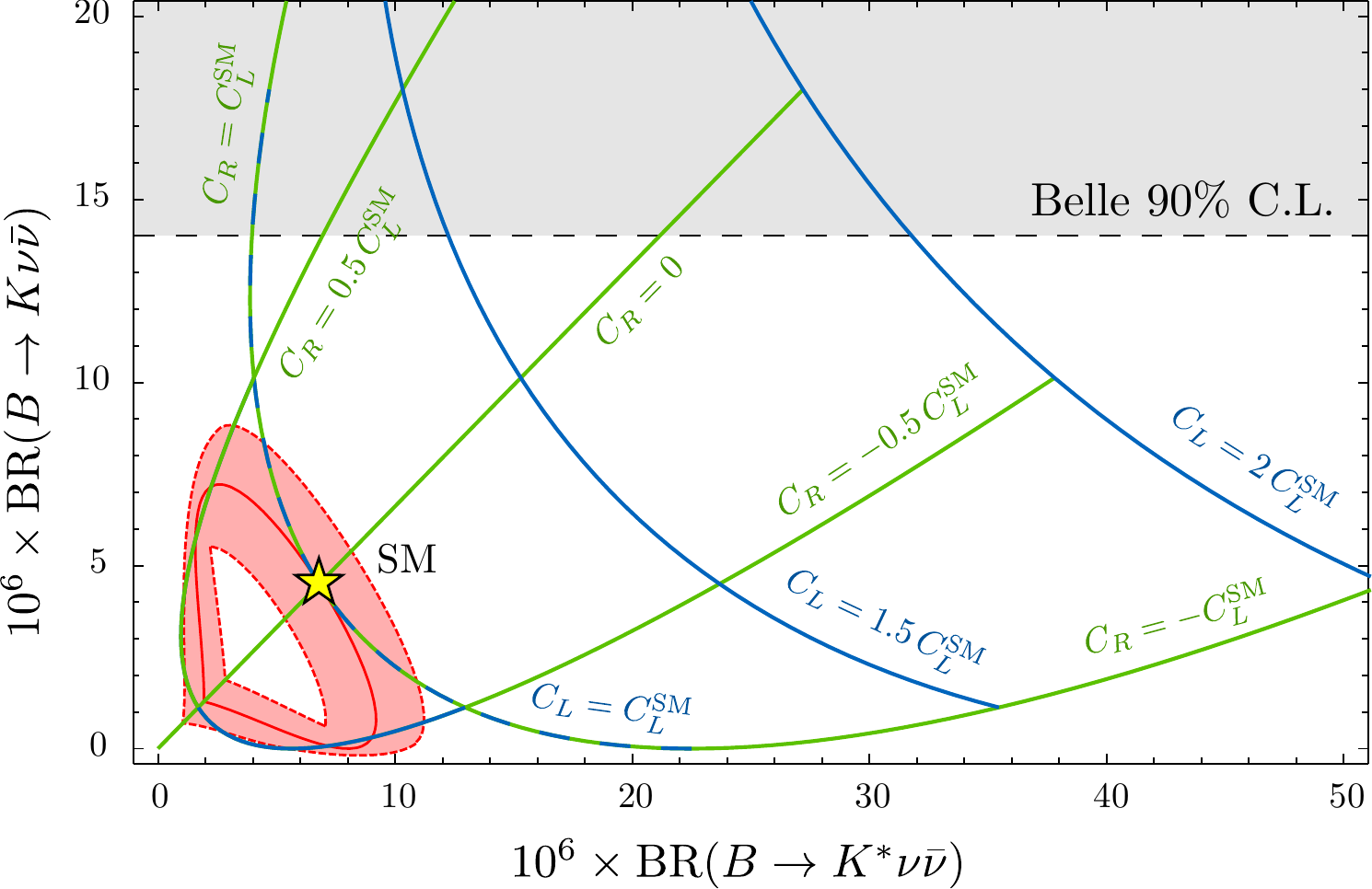}
 \caption{Correlation between the branching ratios of $B \to K^* \nu\bar\nu$ and $B \to K \nu\bar\nu$ depending on the value of the Wilson coefficients $C_L^{bs}$ (blue contours) and $C_R^{bs}$ (green contours) in units of the SM value of $C_L^{bs}$. The gray region is experimentally excluded. The red region is allowed by BR($B\to X_s\ell^+\ell^-$) under the assumption of $Z$ penguin dominance as described in the text. The SM point is marked by a star.}
 \label{fig:bsnunu}
\end{figure}

Assuming that NP enters only the left-handed coefficient $C^{bs}_L$, which is non-zero also in the SM, $\eta=0$ and the branching ratios are simply rescaled, leading to the correlation marked by the line labelled $C_R=0$ in figure~\ref{fig:bsnunu}. If, on the other hand, new physics enters in the form of right-handed currents, with $C^{bs}_L$ being at its SM value, one obtains an anticorrelation between the two branching ratios, as shown in the curve labelled $C_L=C_L^\text{SM}$ (see also \cite{Buras:2010pz}). Any observed deviation from these two curves would signal the presence of NP in both left- and right-handed currents.

This shows that a simultaneous measurement of both decays is desirable to identify the chirality structure of any NP contribution and it demonstrates that NP effects can show up in either of the decays even if its counterpart is found to be SM-like.

Finally, making the assumption, along the lines of strategy 1., that NP effects enter the operators in (\ref{eq:ops}) only through modified $Z$ penguins, a ``horizontal'' correlation can be inferred between $b\to s\nu\bar\nu$ and $b\to s\ell^+\ell^-$ decays. In this case, the good agreement of the branching ratio of the inclusive $B\to X_s\ell^+\ell^-$ decay in the low $q^2$ region with its SM prediction leads to a strict bound on the plane of figure~\ref{fig:bsnunu}, excluding everything outside of the red band. However, this assumption would be very strong, since $b\to s\ell^+\ell^-$ transitions are strongly affected by the magnetic penguin operator $Q_7^{bs}$, so large effects in $b\to s\nu\bar\nu$ are still possible in general.

\section{Final comments}

Rare decays and correlations between them are important tools, complementary to the high-energy frontier, to understand the flavour structure of physics beyond the SM. One message of this talk is that, as long as we are ignorant about the nature of the NP, we should keep an open mind for unexpected effects in rare decays. For example, decays that may seem less promising in MFV -- like $B_d\to\mu^+\mu^-$ compared to $B_s\to\mu^+\mu^-$,  $K_L\to\pi^0\nu\bar\nu$ compared to $K^+\to\pi^+\nu\bar\nu$ or $B\to K^*\nu\bar\nu$ compared to $B\to K\nu\bar\nu$ -- might well turn out to be more interesting than their counterparts.

Finally, one should not forget that equally interesting correlations, that are beyond the scope of this talk, can arise between $\Delta F=1$ and $\Delta F=2$ transitions, among $\Delta F=2$ observables or with rare $D$ decays. 
The coming decade will likely reveal to which point all these correlations will collapse.

\Acknowledgements
I thank Christoph Promberger for sharing data on the SM4, Wolfgang Altmannshofer for data on flavour models, Andrzej Buras for useful discussions and Martin Gorbahn for comments on the manuscript. Finally, I wish to thank the working group III convenors for their invitation and the organizers of CKM2010 for their splendid organization.

This work was supported by the DFG cluster of excellence ``Origin and Structure of the Universe''
and by the EU ITN ``Unification in the LHC Era'', contract PITN-GA-2009-237920 (UNILHC).

\end{document}

%% file: econfmacros.tex
\def\beq{\begin{equation}}
\def\eeq#1{\label{#1}\end{equation}}
\def\eeqn{\end{equation}}

\def\beqa{\begin{eqnarray}}
\def\eeqa#1{\label{#1}\end{eqnarray}}
\def\eeqan{\end{eqnarray}}

\let\bar=\overbar

\def\Dslash{\not{\hbox{\kern-4pt $D$}}}
\def\dslash{\not{\hbox{\kern-2pt $\del$}}}

\def\msb{{\bar{\ssstyle M \kern -1pt S}}}

